\documentclass[prd,showpacs,showkeys,twocolumn]{revtex4}
\usepackage{bm}
\usepackage{amssymb,amsmath,amsthm,amssymb}
\begin{document}
\title[Solution generating theorems for the TOV equation]
{Solution generating theorems for the TOV equation}

\author{Petarpa Boonserm}
\email{Petarpa.Boonserm@mcs.vuw.ac.nz}
\affiliation{School of Mathematics, Statistics, and Computer Science, 
Victoria University of Wellington, PO Box 600, Wellington, New Zealand\\}
\author{Matt Visser}
\email{matt.visser@vuw.ac.nz}
\affiliation{School of Mathematics, Statistics,  and Computer Science, 
Victoria University of Wellington, PO Box 600, Wellington, New Zealand\\}
\author{Silke Weinfurtner}
\email{silke.weinfurtner@vuw.ac.nz}
\affiliation{School of Mathematics, Statistics,  and Computer Science, 
Victoria University of Wellington, PO Box 600, Wellington, New Zealand\\}
\date{5 June 2007;
\LaTeX-ed \today }
\begin{abstract}
  The Tolman--Oppenheimer--Volkov [TOV] equation constrains the
  internal structure of general relativistic static perfect fluid
  spheres. We develop several ``solution generating'' theorems for the
  TOV equation, whereby any given solution can be ``deformed'' into a new
  solution.  Because the theorems we develop work directly in terms of
  the physical observables --- pressure profile and density profile
  --- it is relatively easy to check the density and pressure profiles
  for physical reasonableness. This work complements our previous
  article [Phys.~Rev.~{\bf D71} (2005) 124307; gr-qc/0503007] wherein
  a similar ``algorithmic'' analysis of the general relativistic
  static perfect fluid sphere was presented in terms of the spacetime
  geometry --- in the present analysis the pressure and density are
  primary and the spacetime geometry is secondary. In particular, our
  ``deformed'' solutions to the TOV equation are conveniently
  parameterized in terms of $\delta \rho_c$ and $\delta p_c$, the finite
  \emph{shift} in the central density and central pressure. We conclude by presenting a new physical and mathematical interpretation for the TOV equation --- as an integrability condition on the density and pressure profiles.

\end{abstract}
\maketitle
\def\d{{\mathrm{d}}}
\newtheorem*{theorem}{Theorem}
\def\sint{{\hbox{\small $\displaystyle\int$}}} 
\def\dint{{\hbox{$\displaystyle\int$}}} 
\def\lint{\hbox{\Large $\displaystyle\int$}} 
\def\hint{\hbox{\Huge $\displaystyle\int$}}  
\def\sign{{\mathrm{sign}}}

\section{Introduction}

The general relativistic static perfect fluid sphere has a long and
venerable history that nevertheless continues to provide significant
surprises~\cite{Delgaty, Skea, exact, Petarpa, Martin1,
  Rahman, Lake, Buchdahl, Bondi, Wyman, Hojman-et-al, Martin0}. As emphasised
in the review article by Delgaty and Lake~\cite{Delgaty}, while it is
often relatively easy to write down explicit spacetime metrics that
solve the differential equation corresponding to a relativistic static
perfect fluid sphere, it is often much more difficult to check whether
the corresponding pressure profile and density profile is ``physically
reasonable'' --- indeed there is often some confusion as to what the
phrase ``physically reasonable'' might entail. 

This problem has if
anything become even more acute with the recent introduction of
``algorithmic'' techniques that are capable of generating, and in
principle classifying, \emph{all} perfect fluid spheres~\cite{Petarpa,
  Martin1, Rahman, Lake}. In view of this difficulty we have
reformulated those ``algorithmic'' techniques in terms of several new
``solution generating theorems'' that apply directly to the TOV equation itself
--- and so directly give information about the pressure profile and
density profile.
Specifically, let us consider the well-known TOV system of equations, whose
derivation is now a common textbook
exercise~\cite{Wald,Misner}\,:
\begin{eqnarray}
{\d p(r)\over \d r} &=&
- { [\rho(r)+p(r)]\;[m(r)+ 4\pi p(r)\, r^3]
\over 
r^2 [1-2m(r)/r]};
\\
{ \d m(r)\over \d r} &=& 4\pi \,\rho(r) \, r^2.
\end{eqnarray}
We shall adopt units where $G=c=1$, so that both $m/r$ and $p\, r^2$ are
``dimensionless''.

Our basic strategy will be to assume that somehow we have obtained, or
been given, some specific ``physically reasonable'' solution of the
TOV equation in terms of the ``initial'' profiles $p_0(r)$ and $\rho_0(r)$, and
then to ask how these initial profiles can be \emph{deformed} while still
continuing to satisfy the TOV equation. The output from the analysis will be
several single-parameter, or sometimes multi-parameter, distorted profiles
$p_\Sigma(r)$ and $\rho_\Sigma(r)$ that continue to solve the TOV
equation for arbitrary values of the parameters $\Sigma$, along with
some results regarding the ``physical reasonableness'' thereof. 

We
shall use notation such as $p_{0,c}$ and $p_c$, and $\rho_{0,c}$ and
$\rho_c$, to denote the central pressure, and central density, before
and after applying the distortion. 
Viewing the general relativistic static perfect fluid sphere as a
first-approximation to a relativistic star, (though the discussion could just as easily be applied to a perfect fluid planet),  we will concentrate on the
region deep inside the stellar core, and specifically on the
regularity conditions at the centre of the star, asking that the
pressure and density remain positive and non-singular there. Observe
that in seeking solutions of the TOV equation we are effectively only demanding
pressure isotropy --- we make no claims as to constraints on, or even
the existence of, an equation of state.

We finally conclude by presenting a new physical and mathematical interpretation for the TOV equation --- as an integrability condition on the density and pressure profiles that allows us to explicitly write down the spacetime metric for a relativistic perfect fluid sphere in terms of an iterated integral involving the density and the pressure profiles.

\section{Locating the surface}

As usual, we shall take the
location of the first pressure zero $p(r_s)=0$ as defining the surface of the
star, $r_s$, though we should note a number of caveats: 

\begin{enumerate}

\item  The ``star'' might not have a well-defined surface if $p(r)\neq0$ for all $r$ --- this corresponds to a ``cosmological'' configuration where a localized star-like object is surrounded by an infinite diffuse ``atmosphere'' which merges continuously into some asymptotic limit. While physically there is nothing necessarily wrong with such a configuration, they are certainly non-traditional~\cite{Delgaty}. 

\item If the ``star'' is embedded in an asymptotically de Sitter or anti-de Sitter spacetime then the surface of the star is located by asking that the \emph{material} pressure be zero. The \emph{total} pressure at the surface is then determined by the cosmological constant, so the total pressure is nonzero at the surface: $p_\mathrm{total}(r_s) = p_\Lambda = - \rho_\Lambda \neq 0$.  Though at first this may seem a little alarming, at worst this situation introduces some minor modifications of the standard analysis. 

\item The situation we shall implicitly be most interested in is where the spacetime is asymptotically flat and the surface occurs at some finite radius $r_s$. To avoid needing too many special case technical analyses we shall always take our initial density and pressure profiles $\rho_0(r)$ and $p_0(r)$ to mathematically extend to arbitrary $r$, including beyond $r_s$.  When we specifically want the physical density and pressure profiles we shall explicitly truncate them at the surface  --- for instance (in the pure vacuum case) by setting 
\begin{equation}
m_\mathrm{truncated}(r>r_s) = m(r_s) = M,
\end{equation}
and 
\begin{equation}
\rho_\mathrm{truncated}(r>r_s)=0;
\end{equation}
\begin{equation}
p_\mathrm{truncated}(r>r_s)=0.
\end{equation}
These truncated profiles are certainly a solution to the TOV equation,  and amount (in accordance with Birkhoff's theorem) to continuously joining the surface of  the ``star'' to a portion of the exterior Schwarzschild geometry. 

\item If one wishes to embed a finite size star in  de~Sitter or anti-de~Sitter spacetime then the relevant truncation conditions are
 \begin{equation}
m_\mathrm{truncated}(r>r_s) = m(r_s) + {4\pi\over3} \rho_\Lambda (r^3-r_s^3),
\end{equation}
and 
\begin{equation}
\rho_\mathrm{truncated}(r>r_s)=\rho_\Lambda;
\end{equation}
\vskip 0pt
\begin{equation}
p_\mathrm{truncated}(r>r_s)=p_\Lambda = - \rho_\Lambda.
\end{equation}
These truncated profiles are again certainly a solution to the TOV equation,  and amount (in accordance with a suitable generalization of Birkhoff's theorem) to continuously joining the surface of  the ``star'' to a portion of the Schwarzschild--de~Sitter [Kottler] geometry. 

\end{enumerate}

We shall now use this general framework to develop several mathematical theorems regarding the TOV equation, and subsequently study their physical implications.

\begin{widetext}

\section{The TOV equation as a Riccati equation}

Viewed as a differential equation for $p(r)$, with $m(r)$ [and hence
$\rho(r)$] specified, the TOV equation is a specific example of a
Riccati equation, for which the number of useful solution techniques
is rather limited. (See for instance, Bender \& Orszag~\cite{Bender},
or Polyanin and Zaitsev~\cite{Polyanin}. The most important key results are collected in the appendix to this
article.) Using these standard references it is relatively easy to show
the following.

\begin{theorem}[{\bf P1}]
  Let $p_0(r)$ and $\rho_0(r)$ solve the TOV equation, and hold
  $\displaystyle m_0(r) = 4\pi \int \rho_0(r) \, r^2 \, \d r$ fixed. \\
  Define an auxiliary function $g_0(r)$ by
\begin{equation}
\label{g0_definition}
g_0(r) = {m_0(r)+4\pi p_0(r) \,r^3\over r^2[1-2m_0(r)/r]}.
\end{equation}
Then the general solution to the TOV equation is $p(r) = p_0(r)+\delta
p(r)$ where
\begin{equation}
\label{E:P1}
\delta p(r) =
{
{\displaystyle
\delta p_c \;\sqrt{1-2m_0/r} \; 
\exp\left\{ - 2 \int_{0}^r g_0 \,\d r \right\} 
}
\over
{\displaystyle
1 + 4\pi \delta p_c \int_{0}^r {\displaystyle{1\over \sqrt{1-2m_0/r}}}  
\exp\left\{ -2\int_{0}^r g_0 \,\d r \right\} r \, \d r
}
},
\end{equation}
and where $\delta p_c$ is the shift in the central pressure.
\end{theorem}

\begin{proof}
By using equation \eqref{E:R1} from the appendix we know that for any arbitrary real parameter $k$ the quantity 
\begin{equation}
\delta p(r) = { 
{\displaystyle
k \exp\left[ - \int {m_0+4\pi \rho_0 r^3 
+ 8\pi \, p_0 r^3\over r^2(1-2m_0/r)} \; \d r\right]
}
\over
{\displaystyle 
1+ 4\pi k  \dint {r\over 1-2m_0/r} 
\exp\left[ - \dint {m_0+4\pi \rho_0  r^3 
+ 8\pi \, p_0  r^3\over r^2(1-2m_0/r)} \d r\right]  \d r 
}
}
\end{equation}
is such that $p(r)=p_0(r)+\delta p(r)$ also solves the TOV equation. But by using an integration by parts one can simplify this to establish
\begin{eqnarray}
\nonumber
\exp\left[ -\int {m_0+4\pi \, \rho_0 \, r^3 
+ 8\pi \, p_0 \, r^3\over r^2(1-2m_0/r)} \; \d r\right]
&=& {\sqrt{1-2m_0/r}} \; 
\exp\left[ -2 \int {m_0+4\pi\, p_0 \,r^3\over r^2(1-2m_0/r)} \; \d r\right] 
\\
&=&  \sqrt{1-2m_0/r}  \; \exp\left[-2\int g_0 \; \d r\right].
\end{eqnarray}
Thus
\begin{equation}
\delta p(r) = { 
{\displaystyle
k \sqrt{1-2m_0/r}  \; \exp\left[-2\int g_0 \; \d r\right]
}
\over
{\displaystyle 
1+ 4\pi k   \int {\displaystyle{1\over \sqrt{1-2m_0/r}}}  
\exp\left\{ -2\int g_0 \,\d r \right\} r \, \d r
}
}.
\end{equation}
Finally, one fixes the constant $k$ by first choosing the limits of integration to be $0$ to $r$, and then applying the boundary condition $p(0)=p_c$
at $r=0$ to deduce $k=\delta p_c$. Thus
\begin{equation}
\delta p(r) =
{
{\displaystyle
\delta p_c \;\sqrt{1-2m_0/r} \; 
\exp\left\{ - 2 \int_{0}^r g_0 \,\d r \right\} 
}
\over
{\displaystyle
1 + 4\pi \delta p_c \int_{0}^r {\displaystyle{1\over \sqrt{1-2m_0/r}}}  
\exp\left\{ -2\int_{0}^r g_0 \,\d r \right\} r \, \d r
}
}
\end{equation}
as claimed.
\end{proof}

\end{widetext}

\subsection{Physically reasonable centre} 

If the original $p_0(r)$ and
$m_0(r)$ are ``physically reasonable'', in particular if the central
pressure $p_{0,c}$ and density $\rho_{0,c}$ are finite and positive,
then $m_0(r)= O(r^3)$, so $m_0/r = O(r^2)$, and $g_0(r)=O(r)$. Thus
all the integrals used above are finite and well-behaved (convergent)
at the centre of the star.  Consequently, the new central pressure
will likewise be finite and well behaved; and will even be positive so
long as $\delta p_c > - p_{0,c}$.  So the central region of the new
stellar configuration will automatically be ``physically reasonable''.
Since within the context of this theorem the density profile is
unaffected, we also know that no horizon will be generated if we start
from a horizon-free initial state. (Horizon formation is impossible in the current situation: If it were to
happen, it would be signaled by the existence of an $r_H$ such that
$2m(r_H)/r_H=1$, but by hypothesis $\rho(r)$, and hence $m(r)$, are held fixed. So if there is no horizon in the original configuration, one cannot generate a horizon by applying the current theorem.)

\noindent
\emph{Comment:} With hindsight one can view this theorem {\bf P1}
as a consequence of theorem {\bf T2} of reference~\cite{Petarpa}. Some
tedious algebraic manipulations will actually show that they are
identical, up to a specific regularity choice. (In the
notation of~\cite{Petarpa}, one must choose $\sigma\neq 0$.)  This regularity condition is now \emph{forced}
upon us because we want the pressure and density at the center of the
star to remain finite.  

Equivalently, if the original $p_0(r)$ and
$m_0(r)$ are ``physically reasonable'', then (provided the relevant
integrals are chosen to run from $0$ to $r$), the $\sigma =0$ sub-case
of theorem {\bf T2} of reference~\cite{Petarpa} yields a spacetime
geometry for which $g_{tt}\to0$ at the origin, which we deem to
\emph{not} be physically reasonable.  The key difference in the
physics of the current article is that we now have a physics reason
for restricting the solutions generated by theorem {\bf T2} of
reference~\cite{Petarpa}, and now can thereby derive an explicit
theorem directly in terms of the shift in the pressure profile.

\subsection{Central redshift}

If one writes the spacetime geometry in the
form
\begin{equation}
\d s^2 = - \zeta_0(r)^2 \, \d t^2 
+ {\d r^2\over 1-2m_0(r)/r} + r^2 \; \d^2 \Omega,
\end{equation}
then the Einstein equations
imply (see, for instance,~\cite{Petarpa, Wald, Misner})
\begin{equation}
\label{zeta_result}
\zeta_0(r) = \exp\left\{ \int g_0(r) \d r \right\}.
\end{equation}
It is important to realise that with current conventions $g_0(r)$ is
\emph{defined} by equation (\ref{g0_definition}), and that equation
(\ref{zeta_result}) is a \emph{result} of the Einstein equations.

In particular, throughout the star
\begin{equation}
\label{zeta_result2}
\zeta_0(r) = \zeta_{0,c} \; \exp\left\{ \int_0^{r} g_0(r) \d r \right\}, \quad (r\leq r_s),
\end{equation}
and 
at the surface of the star
\begin{equation}
\zeta_0(r_s) = \zeta_{0,c} \; \exp\left\{ \int_0^{r_s} g_0(r) \d r \right\},
\end{equation}
where $\zeta_{0,c}$ denotes the value of $\zeta_0(r)$ at the centre of
the fluid sphere --- which is effectively the central redshift of the fluid sphere.  
In fact for an idealized freely propagating particle emitted from the centre
\begin{equation}
\zeta_{0,c} = {1\over 1+ z_c}.
\end{equation}
If one truncates the matter profile at $r_s$, and 
and normalizes using in the usual way using $\zeta_\mathrm{truncated}(\infty)=1$, then because of the Birkhoff theorem the geometry exterior to the star is Schwarzschild, and so one also has:
\begin{equation}
\zeta_0(r_s)  = \sqrt{1-2M/r_s}.
\end{equation}
Consequently
\begin{equation}
  \zeta_{0,c} =  \sqrt{1-2M/r_s} \exp\left\{ - \int_0^{r_s} g_0(r) \d r \right\}.
\end{equation}
Note that choosing the normalization $\zeta_\mathrm{truncated}(\infty)=1$ means that we are choosing the $t$ coordinate to be proper time as measured at spatial infinity --- which in general is well outside the surface $r_s$ of the star. 
For $r\leq r_s$ one could if one chooses rewrite equation (\ref{zeta_result2}) in the completely equivalent  form
\begin{equation}
\zeta_0(r) = 
\sqrt{1-2M/r_s} \; \exp\left\{ -\int_r^{r_s} g_0(r) \d r \right\},  \quad (r\leq r_s).
\end{equation}

These considerations now allow one to rewrite the result of theorem {\bf P1} as:
\begin{equation}
\label{E:P1-2}
\delta p(r) = {\delta p_c \;\zeta_{0,c}^2 \;  \sqrt{1-2m_0(r)/r} 
\over
{\displaystyle
 \zeta_0(r)^2
\left[1 + 4\pi \delta p_c \; \zeta_{0,c}^2 \;
\int_{0}^r {  r \, \d r \over\sqrt{1-2m_0/r} \; \zeta_0(r)^2} \right]
}
}.
\end{equation}
This has allowed us to eliminate the inner integral (involving $g_0$)
in terms of the $g_{tt}$ metric component.

\subsection{Shift in the surface}

From either equation (\ref{E:P1}) or (\ref{E:P1-2}), since all the other terms are positive, we see that the \emph{sign} of $\delta p(r)$ is the same as the \emph{sign} of $\delta p_c$, and so the same as the \emph{sign} of the shift in the surface radius $\delta r_s$. This is most obvious for $\delta p_c>0$ where the positivity of $\delta p(r)$ implies that the surface definitely moves outwards. For small $\delta p_c<0$ the negativity of $\delta p(r)$ implies that the surface moves inwards --- the tricky point is that for negative $\delta p_c$ one could in principle (mathematically) encounter a pole at some finite $r_\mathrm{pole}$, but this would then imply $p(r_\mathrm{pole}) = -\infty$, so the putative pole cannot occur inside the star --- it is at worst a mathematical artifact located at $r_\mathrm{pole} > r_s$, outside the physical surface of the star. All in all,
\begin{equation}
\sign[\delta r_s] = \sign[\delta p_c].
\end{equation}

We can say a little more if we restrict attention to infinitesimal shifts in the central pressure $\delta p_c$ and take derivatives. The surface of the star is defined by
\begin{equation}
p(p_c, r_s(p_c))=0,
\end{equation}
where
\begin{equation}
p(p_c, r)=p_0(r) + \delta p(\delta p_c,r).
\end{equation}
Thus
\begin{equation}
\left.{\partial p\over\partial p_c}\right|_{r_s,p_{0,c}} + 
\left.{\partial p\over\partial r}\right|_{r_s,p_{0,c}} \; 
\left.{\partial r_s\over \partial p_c} \right|_{p_{0,c}} = 0.
\end{equation}
That is
\begin{equation}
\label{E:shift-in-surface}
\left.{\partial r_s\over \partial p_c} \right|_{p_{0,c}} = 
- 
{
\left.{\partial p/\partial p_c}\right|_{r_s,p_{0,c}}
\over
\left.{\partial p/\partial r}\right|_{r_s,p_{0,c}} 
}.
\end{equation}
But now
\begin{eqnarray}
\left.{\partial p\over\partial p_c}\right|_{r_s,p_{0,c}}  
&=& \left.{\partial (\delta p)\over\partial (\delta p_c)}\right|_{r_s,\delta p_c=0},
\\
&=& \sqrt{1-2M/r_c} \; \exp\left\{ -2 \int_0^{r_s} g_0\, \d r\right\}, \;\;\;
\\
&=& \zeta_{0,c}^2/ \sqrt{1-2M/r_s}.
\end{eqnarray}
Furthermore, from the TOV equation
\begin{equation}
\label{E:TOV-at-surface}
\left.{\partial p\over\partial r}\right|_{r_s,p_{0,c}}  = - {\rho_0(r_s) M\over r_s^2 [1-2M/r_s]}.
\end{equation}
Combining, we see
\begin{equation}
\left.{\partial r_s\over \partial p_c} \right|_{p_{0,c}} = 
{\zeta_{0,c}^2\;r_s^2\;\sqrt{1-2M/r_s}\over \rho_0(r_s)\; M}.
\end{equation}
All terms appearing above are manifestly positive.

\subsection{Shift in the compactness} 

If we now consider the ``compactness'', defined as 
\begin{equation}
\chi = {2M\over r_s} = {2m(r_s)\over r_s},
\end{equation}
then, since $m(r)=m_0(r)$ is invariant under the current hypotheses, 
\begin{equation}
\delta \chi = {8\pi\over 3} \; \left\{ 3 \rho_0(r_s) -  \bar \rho_0(r_s) \right\} \; r_s \; \delta r_s 
+ \mathcal{O}([\delta r_s]^2),
\end{equation}
where $\bar \rho_0(r_s)$ is the ``average density'' defined by $m_0(r_s)/({\small{4\pi\over3}} r_s^3)$. Unfortunately, while (as we have seen above) the \emph{sign} of $\delta r_s$ is controlled, there is in general no \emph{a priori} constraint on the \emph{sign} of $3\rho_0(r_s)-\bar\rho_0(r_s)$, so we cannot (without further assumptions) constrain the \emph{sign} of the shift in compactness. This means that theorem {\bf P1} by itself will not tell you if you are approaching or receding from the Buchdahl--Bondi bound $\chi_\mathrm{maximum}=8/9$. The best we can say is
\begin{equation}
\sign[\delta\chi] = \sign[3\rho_0(r_s)-\bar\rho_0(r_s)] \times \sign[\delta r_s].
\end{equation}
Note that the function $\left\{ 3 \rho_0(r) -  \bar \rho_0(r) \right\}= \left\{ 3 \rho(r) -  \bar \rho(r) \right\}$ remains invariant under the conditions of theorem {\bf P1}.  Whether or not this quantity is positive or negative  depends on one's \emph{a priori} arbitrary choice of the initial solution profile $\{\rho_0(r),\,p_0(r)\}$ used as input into the theorem.
If we now consider infinitesimal shifts in the central pressure we see
\begin{equation}
{\partial \chi\over\partial r_s} = {8\pi\over 3} \; \left\{ 3 \rho_0(r_s) -  \bar \rho_0(r_s) \right\} \; r_s,
\end{equation}
and via the chain rule
\begin{equation}
{\partial \chi\over\partial p_c} = {8\pi\over 3} \; 
{\left\{ 3 \rho_0(r_s) -  \bar \rho_0(r_s) \right\} \; r_s^3 \; \zeta_{0,c}^2 \; \sqrt{1-2M/r_s} 
\over
\rho_0(r_s)\; M}.
\end{equation}

\begin{widetext}
\subsection{Shift in the spacetime geometry} 

One might also reasonably ask what happens to the spacetime metric
itself? As regards the radial $g_{rr}$ part of the spacetime metric,
$m_0(r)$ is by construction fixed, and so $g_{rr}$ is unaffected.  To
calculate the ``deformed'' value of the the $g_{tt}= - \zeta(r)^2$ component we note that from
equation \eqref{g0_definition} we have
\begin{equation}
\delta g(r) = {4\pi\;\delta p(r)\; r\over1-2m_0(r)/r},
\end{equation}
whence
\begin{equation}
\delta g(r) = 
{
{\displaystyle 
{4\pi\;\delta p_c\; r\over \sqrt{1-2m_0/r}} 
\exp\left\{ - 2 \int_{0}^r g_0 \,\d r \right\} 
}
\over
{\displaystyle
1 + 4\pi \delta p_c \int_{0}^r {\displaystyle{r \, \d r \over \sqrt{1-2m_0/r}}}  
\exp\left\{ -2\int_{0}^r g_0 \,\d r \right\}
}   
}.
\end{equation}
That is
\begin{equation}
\delta g(r) =  {\d\over\d r} \ln \left\{
{\displaystyle
1 + 4\pi \delta p_c \int_{0}^r {\displaystyle{r \, \d r \over \sqrt{1-2m_0/r}}}  
\exp\left\{ -2\int_{0}^r g_0 \,\d r \right\}
}   
\right\}.
\end{equation}

Note that the Einstein equations imply, \emph{vide} \eqref{zeta_result}--\eqref{zeta_result2},
\begin{equation}
\ln(\zeta/\zeta_0) 
= \ln(\zeta_c/\zeta_{0,c}) + \int_0^r \delta g\; \d r.
\end{equation}
Integrating, 
\begin{eqnarray}
&&\!\!\!\!\!\!
\ln(\zeta/\zeta_0) =  \ln(\zeta_c/\zeta_{0,c}) 
 + \ln\left\{
{
1 + 4\pi \delta p_c \int_{0}^r {
{r \, \d r \over \sqrt{1-2m_0/r}}}  
\exp\left\{ -2\int_{0}^r g_0 \,\d r \right\}
}   
\right\}.
\end{eqnarray}

The only tricky part of the computation lies in getting the overall
normalization correct. Since $g_0(r)$ is defined by
\eqref{g0_definition} directly in terms of mass and pressure profiles,
the above formula provides a completely explicit formula for the
(multiplicative) shift in $\zeta$, and so for the (multiplicative)
shift in $g_{tt}$.  Explicitly,
\begin{eqnarray}
{\zeta(r)\over\zeta_c} &=& {\zeta_0(r)\over\zeta_{0,c}} 
\times
\left\{ 
1 + 4\pi \delta p_c \; \zeta_{0,c}^2 \;
\int_{0}^r { r \, \d r \over\sqrt{1-2m_0/r} \; \zeta_0(r)^2}
\right\}.
\end{eqnarray}

In view of these formulae for $\zeta(r)$ we can rewrite the shift in
pressure in the more compact manner
\begin{equation}
\delta p(r) = 
{
{\displaystyle
\delta p_c \;\sqrt{1-2m_0/r} \; 
\zeta_{0,c} \; \zeta_c
}
\over
\zeta_0(r) \; \zeta(r)
}.
\end{equation}
Though significantly more compact, the trade-off in this formula is
that $\zeta(r)$ and $\zeta_0(r)$ are themselves quite complicated
functions of the mass and pressure profiles.

\noindent\emph{Normalization of theorem {\bf P1} versus theorem {\bf T2}:} 
With the above normalization results now in hand, we see that theorem
{\bf P1} corresponds to theorem {\bf T2} of \cite{Petarpa} with the integrals running
from $0$ to $r$, with $\sigma\to \zeta_c/\zeta_{0,c}$, while the other
parameter $\epsilon$ occurring in {\bf T2} can be physically
identified by taking $\epsilon/\sigma \to 4\pi\;\delta p_c \;
\zeta_{0,c}^2$.

\subsection{Two additional theorems} 

In the specific case considered in this section, where $m(r)$
is held fixed, one can obtain two additional related theorems by using equations \eqref{E:R2} and
\eqref{E:R3} of the appendix.  Specifically:

\begin{theorem}[{\bf P1b}]
  For a fixed specification of $\rho_0(r)$, let $p_1(r)$ and $p_2(r)$
  be two distinct solutions of the TOV equation. Then the general
  solution (for $\lambda$ an arbitrary real parameter) is

\begin{equation}
p(r) =
{
{\displaystyle
\lambda  
\exp\left\{-\int_0^r {4\pi p_1\over 1-2m_0/r} \; r \,\d r \right\}  p_1(r) 
+ (1-\lambda) 
\exp\left\{-\int_0^r {4\pi p_2\over 1-2m_0/r} \; r \,\d r \right\}  p_2(r)
}
\over
{\displaystyle
\lambda 
\exp\left\{-\int_0^r  {4\pi p_1\over 1-2m_0/r} \;r \, \d r \right\}  
+ (1-\lambda) 
\exp\left\{-\int_0^r  {4\pi p_2\over 1-2m_0/r} \;r \,\d r \right\}
}
}.
\end{equation}
The central pressure is
\begin{equation}
p_c =
\lambda \; p_{c,1} + (1-\lambda) \; p_{c,2}.
\end{equation}
\end{theorem}
\noindent
Note that this theorem no longer requires any nested integrations, at
the cost of needing two specific solutions as input. A minor technical issue is that $p_1(r)$ and $p_2(r)$ have to solve exactly the \emph{same} Riccati equation over the \emph{entire} domain of interest --- this leads to technical problems if we truncate the density and pressure profiles at the surface of the star. We will either be forced to work on the intersection of the truncation regions, that is $r\in[0,\mathrm{min}\{r_{s,1},r_{s,2}\}]$, or work with un-truncated pressure profiles $p_1(r)$ and $p_2(r)$ that have been extended into the region where pressure is permitted to become negative. 

\begin{theorem}[{\bf P1c}]
  For a fixed specification of $\rho_0(r)$, let $p_1(r)$, $p_2(r)$,
  and $p_3(r)$ be three distinct solutions of the TOV equation. Then
  the general solution (for $\lambda$ an arbitrary real parameter) is
\begin{equation}
p(r) = { \lambda \;p_1(r)\; [p_3(r)-p_2(r)] + (1-\lambda) \;p_2(r)\; [p_3(r)-p_1(r)]
\over
\lambda \;[p_3(r)-p_2(r)] + (1-\lambda) \; [p_3(r)-p_1(r)]}.
\end{equation}
The central pressure is then
\begin{equation}
p_c = 
{ \lambda \;p_{c,1}\; [p_{c,3}-p_{c,2}] + (1-\lambda) \;p_{c,2}\; [p_{c,3}-p_{c,1}]
\over
\lambda \;[p_{c,3}-p_{c,2}] + (1-\lambda) \; [p_{c,3}-p_{c,1}]}.
\end{equation}
\end{theorem}
\noindent
Note that this last theorem no longer requires any
integrations whatsoever, though now at the cost of needing three specific solutions as
input. As in the preceding theorem,  $p_1(r)$,  $p_2(r)$, and $p_3(r)$ have to solve exactly the \emph{same} Riccati equation over the \emph{entire} domain of interest --- for stars with a well defined surface we will either be forced to work on the intersection of the truncation regions, that is $r\in[0,\mathrm{min}\{r_{s,1},r_{s,2},r_{s,3}\}]$, or work with un-truncated pressure profiles $p_1(r)$, $p_2(r)$, and $p_3(r)$ that have been extended into the region where pressure is permitted to become negative. 

\bigskip
\end{widetext}
\smallskip

\section{Correlated changes in density and pressure}

The second main theorem we develop can be obtained by looking for \emph{correlated}
changes in the mass and pressure profiles.

\begin{theorem}[{\bf P2}]
  Let $p_0(r)$ and $\rho_0(r)$ solve the TOV equation, and hold $g_0$
  fixed, in the sense that
\begin{equation}
g_0(r) = {m_0(r)+4\pi p_0(r) r^3\over r^2[1-2m_0(r)/r]} 
=  {m(r)+4\pi p(r) r^3\over r^2[1-2m(r)/r]}.
\end{equation}
Then the general solution to the TOV equation is given by $p(r) =
p_0(r)+\delta p(r)$ and $m(r) = m_0(r)+\delta m(r)$ where
\begin{equation}
\delta m(r) =   {4\pi r^3 \; \delta \rho_c  \over 3\; [1+r\;g_0]^2} \; 
\exp\left\{  2 \int_0^r g_0 \; {1-r\,g_0\over 1+r\,g_0} \; \d r \right\},
\end{equation}
and 
\begin{equation}
\delta p(r) =  - { \delta m\over 4\pi r^3} \; {1+8\pi p_0 r^2 \over 1-2m_0/r}.
\end{equation}
Here $\delta \rho_c$ is the shift in the central density.

Explicitly combining these formulae we have
\begin{eqnarray}
\delta p(r) &=&  {\delta p_c \over  [1+r\;g_0]^2} \; {1+8\pi p_0 r^2 \over 1-2m_0/r}\; 
\nonumber
\\
&& \times
\exp\left\{  2 \int_0^r g_0 \; {1-r\,g_0\over 1+r\,g_0} \; \d r \right\}.
\end{eqnarray}

\end{theorem}

\begin{proof}
Since
\begin{equation}
m(r) = { g_0(r) r^2 - 4\pi p(r) r^3\over 1 + 2 r g_0(r)},
\end{equation}
we have in particular
\begin{equation}
\delta m(r) = - {4\pi\, \delta p(r) \,r^3\over 1 + 2 r g_0(r)} 
= - 4\pi \delta p(r) r^3\; \left\{ { 1-2m_0/r\over 1+8\pi p_0 r^2} \right\},
\end{equation}
so that
\begin{equation}
\delta \rho(r) = - {1\over r^2} \; {\d \over \d r} 
\left({\delta p(r) r^3\over 1 + 2 r g_0(r)}\right).
\end{equation}
Now consider the TOV equation, or more precisely, the \emph{change} in
the TOV equation:
\begin{equation}
{\d \delta p(r)\over \d r} = 
-  [\delta \rho(r)+\delta p(r)]\;g_0(r).
\end{equation}
Combining these last two differential equations yields a linear
homogeneous differential equation for $\delta p(r)$, which is easily
integrated. The quoted form of the theorem results after an
integration by parts.
\end{proof}

\subsection{Physically reasonable centre} 

If the original density and
pressure profiles are ``physically reasonable'', then likewise the new
density and pressure profiles will be well behaved --- at least for a
finite region including the origin. In particular, since regularity of
the original solution implies $g_0(r) = O(r)$ at the origin, the only
integral we have to do for this theorem is guaranteed to converge at
the lower limit $r=0$. So the central region of the new stellar
configuration will automatically be ``physically reasonable''.  Note
that for theorem {\bf P2}
\begin{equation}
\delta p_c = - {\delta \rho_c\over 3}.
\end{equation}
If the central pressure increases $\delta p_{c} > 0$, then for the
situation envisaged in this theorem the mass profile always decreases,
implying that an event horizon never forms.

\bigskip
\noindent
\emph{Comment:} With hindsight one can also view this theorem {\bf P2}
as a consequence of theorem {\bf T1} of reference~\cite{Petarpa}. The
key difference now is that we have an explicit statement directly in
terms of the shift in the pressure profile. To see this, write
$g_0=\zeta_0'/\zeta_0$, substitute into {\bf T1} and apply boundary
conditions to deduce $\lambda \to 8\pi\;\delta\rho_c/3$.

\subsection{Shift in the surface} 

Note that under the conditions of theorem {\bf P2} we have
\begin{equation}
\sign[\delta p(r)] = - \sign[\delta m(r)] = -\sign[\delta \rho_c].
\end{equation}
So if the central density goes up, the pressure everywhere goes down, and the radius of the surface of the star (being defined by the first pressure zero) must decrease. That is
\begin{equation}
\sign[\delta r_s] = -\sign[\delta \rho_c] = +\sign[\delta p_c].
\end{equation}

If we now consider infinitesimal changes in central pressure we see that equation (\ref{E:shift-in-surface}) is unaltered. Furthermore, under the hypotheses of theorem {\bf P2} we have
\begin{eqnarray}
\left.{\partial p\over\partial p_c}\right|_{r_s,p_{0,c}}  
&=& \left.{\partial (\delta p)\over\partial (\delta p_c)}\right|_{r_s,\delta p_c=0} 
\\
&=& {1 \over  [1+r_s\;g_0(r_s)]^2 \; [ 1-2M/r_s]}\;  
\nonumber
\\
&& \times \exp\left\{  2 \int_0^{r_s} g_0 \; {1-r\,g_0\over 1+r\,g_0} \; \d r \right\},
\end{eqnarray}
which is manifestly positive. Thanks to the TOV equation we again have equation (\ref{E:TOV-at-surface}), so combining the above leads to
\begin{eqnarray}
\left.{\partial r_s\over \partial p_c} \right|_{p_{0,c}} &=& 
 {r_s^2 \over  [1+r_s\;g_0(r_s)]^2 \;  \rho_0(r_s)\; M}\;   
\nonumber
\\
&& \times \exp\left\{  2 \int_0^{r_s} g_0 \; {1-r\,g_0\over 1+r\,g_0} \; \d r \right\}.
\end{eqnarray}
All terms appearing above are manifestly positive, as was the case for theorem {\bf P1}.

\subsection{Shift in the compactness} 

The compactness $\chi = 2m(r_s)/r_s$ now changes \emph{both} due to changes in $m(r)$ \emph{and} due to changes in $r_s$. For small but finite changes we have
\begin{equation}
\delta\chi = {2 \; \delta m(r_s)\over r} + {8\pi\over 3} \; \left\{ 3 \rho_0(r_s) -  \bar \rho_0(r_s) \right\} \; r_s \; \delta r_s 
+ \dots.
\end{equation}
It is more efficient to re-write the compactness as 
\begin{equation}
\chi(p_c) = {2m(p_c,r_s(p_c))\over r_s(p_c)}
\end{equation}
and then differentiate. We see
\begin{eqnarray}
\left.{\partial \chi\over \partial p_c}\right|_{p_{0,c}} &=& 
{2\over r_s} \left.{\partial m(p_c,r) \over \partial p_c}\right|_{r_s(p_{0,c}), p_{0,c}}
\\
&&
+
 {8\pi\over 3} \; \left\{ 3 \rho_0(r_s) -  \bar \rho_0(r_s) \right\} \; r_s \; 
\left.{\partial r_s(p_c) \over \partial p_c}\right|_{p_{0,c}}.  \;\;\;
\nonumber
\end{eqnarray}
But under the hypotheses of theorem {\bf P2} we have
\begin{eqnarray}
\left.{\partial m(p_c,r_s) \over \partial p_c}\right|_{p_{0,c}} &=&
 -{4\pi r_s^3 \;  \over\; [1+r_s\;g_0]^2} \; 
 \\
&&
\times
\exp\left\{  2 \int_0^{r_s} g_0 \; {1-r\,g_0\over 1+r\,g_0} \; \d r \right\}.
\nonumber
\end{eqnarray}
Now,  combining this with the results of the previous subsection
\begin{eqnarray}
\left.{\partial \chi\over \partial p_c}\right|_{p_{0,c}} &=& 
{8\pi r_s^3\over M [1+r_s\;g_0]^2 } \; 
\exp\left\{  2 \int_0^{r_s} g_0 \; {1-r\,g_0\over 1+r\,g_0} \; \d r \right\}
\nonumber
\\
&&
\times\left\{ 1- {M\over r_s}  - {\bar \rho_0\over3\rho_0(r_s)} \right\}.
\end{eqnarray}
While the first few terms are guaranteed positive, there is no such guarantee as to the sign of the last term. Thus without further hypotheses we again cannot constrain the sign of the shift in the compactness $\chi = 2m(r_s)/r_s$.

\subsection{Shift in the spacetime geometry}
What happens to the spacetime geometry under the application of theorem {\bf P2}? By construction the quantity $g(r)=g_0(r)$ is unaffected, and therefore 
\begin{equation}
{\zeta(r)\over\zeta_c} = {\zeta_0(r)\over\zeta_{0,c}}.
\end{equation}
It is only the $g_{rr}$ component of the spacetime metric that changes significantly:
\begin{equation}
[g_0]_{rr} \to g_{rr} = {1\over [g_0]_{rr}^{-1} - \delta[2m(r)/r]},
\end{equation}
where we already know
\begin{equation}
\delta\left[{2m(r)\over r}\right] =
  {8\pi r^2 \; \delta \rho_c  \over 3\; [1+r\;g_0]^2} \; 
\exp\left\{  2 \int_0^r g_0 \; {1-r\,g_0\over 1+r\,g_0} \; \d r \right\}.
\end{equation}

\section{Combining the previous theorems}

One can obtain additional and more complicated theorems by iteratively applying
theorems {\bf P1} and {\bf P2}, in a manner similar to the discussion
of reference~\cite{Petarpa}. This iteration will yield 2-parameter
generalizations of theorems {\bf P1} and {\bf P2}.

\begin{theorem}[{\bf P3}]
  Let $p_0(r)$ and $\rho_0(r)$ solve the TOV equation. Apply theorem
  {\bf P1} followed by theorem {\bf P2}. Let us define intermediate quantities
\begin{equation}
\delta p_1(r) = {
{\displaystyle
\Delta p \;\sqrt{1-2m_0/r} \; \exp\left\{ - 2 \int_{0}^r g_0 \,\d r \right\} 
}
\over
{\displaystyle
1 + 4\pi \, \Delta p \int_{0}^r {\displaystyle{r \, \d r \over \sqrt{1-2m_0/r}}}  
\exp\left\{ -2\int_{0}^r g_0 \,\d r \right\}
}
},
\end{equation}
and
\begin{equation}
g_1(r) = g_0(r) + {4\pi \, \delta p_1(r)\over1-2m_0/r}.
\end{equation}
Then $p(r) = p_0(r)+\delta p(r)$ and $m(r) = m_0(r)+\delta m(r)$ are
also solutions of the TOV equation, where
\begin{equation}
\delta m(r) =   {4\pi r^3 \; \delta \rho_c  \over 3 \; [1+r\;g_1]^2} \; 
\exp\left\{  2 \int_0^r g_1 \; {1-r\,g_1\over 1+r\,g_1} \; \d r \right\},
\end{equation}
and
\begin{equation}
\delta p(r) = \delta p_1(r) - { \delta m\over 4\pi r^3} \; 
{1+8\pi [p_0+\delta p_1] r^2 \over 1-2m_0/r}.
\end{equation}
Here $\delta \rho_c$ is the shift in the
central density, and the total shift in the central pressure is given by
\begin{equation}
\delta p_c =  \Delta p -{\delta\rho_c\over3}.
\end{equation}
The two parameters $\delta\rho_c$ and $\Delta p$ can be specified
independently.
\end{theorem}

\begin{theorem}[{\bf P4}]
  Let $p_0(r)$ and $\rho_0(r)$ solve the TOV equation. Apply theorem
  {\bf P2} followed by theorem {\bf P1}. Then $p(r) = p_0(r)+\delta
  p(r)$ and $m(r) = m_0(r)+\delta m(r)$ are also solutions of the TOV equation,
  where
\begin{equation}
\delta m(r) =   {4\pi r^3 \; \delta \rho_c  \over 3 \; [1+r\;g_0]^2} \; 
\exp\left\{  2 \int_0^r g_0 \; {1-r\,g_0\over 1+r\,g_0} \; \d r \right\},
\end{equation}
and 
\begin{widetext}
\begin{equation}
\delta p(r) = - { \delta m\over 4\pi r^3} \;\; {1+8\pi p_0 r^2 \over 1-2m_0/r} \;\;
{\Delta p \;\zeta_{0,c}^2  \;\sqrt{1-2[m_0+\delta m]/r} 
\over
{\displaystyle
\zeta_0(r)^2
\left[1 + 4\pi \Delta p \; \zeta_{0,c}^2 \;
\int_{0}^r { r \, \d r  \over\sqrt{1-2[m_0+\delta m]/r} \zeta_0^2(r)} \right]
}
}.
\end{equation}
\end{widetext}
Here $\delta \rho_c$ is the shift in the
central density, and the total shift in the central pressure is given by
\begin{equation}
\delta p_c =  \Delta p -{\delta\rho_c\over3}.
\end{equation}
The two parameters $\delta\rho_c$ and $\Delta p$ can be specified
independently.
\end{theorem}

\noindent
\emph{Comment:} Note that these two theorems provide two-parameter
generalizations of the TOV solution ($\rho_0$, $p_0$) that one starts
from. These theorems are closely related to theorems {\bf T4} and {\bf
  T3} of reference~\cite{Petarpa}.  Technically they are equivalent to
the $\sigma\neq 0$ case of theorems {\bf T4} and {\bf T3} of
reference~\cite{Petarpa}, where $\sigma\neq 0$ is now a physical
restriction we place on the solution by demanding that the density and
pressure be well behaved at the centre of the fluid sphere.

\section{The TOV equation as an Abel equation}

If, on the other hand, we think of the pressure profile $p(r)$ as
fixed, than we can rearrange the TOV system of equations as a
differential equation for $m(r)$, specifically:
\begin{equation}
{\d m(r)\over \d r} = - 4\pi p(r) r^2 
-{4\pi r^4[1-2m(r)/r]\over m(r)+4\pi p(r) r^3} \; {\d p\over\d r}.
\end{equation}
This is now an Abel equation (2nd type, class A).

If one were able to develop a solution generating theorem based on this
equation it would in many ways be the most natural companion to
theorem {\bf P1}.  Unfortunately, despite numerous attempts at
developing solution generating theorems based on this observation, we
have no progress to report. This is ultimately due to the fact that mathematically
Abel equations are considerably more difficult to deal with than
Riccati equations.

\section{The TOV equation as an integrability condition}

We now develop an alternative point of view regarding the physical and mathematical significance of the TOV equation: In any spherically symmetric spacetime there are formally three algebraically independent components of the Einstein equations. Let us now formally solve two of the Einstein equations by building them into the chosen form of the line element, and use the remaining Einstein equation to derive (in the case of perfect fluids) an integrability condition --- this provides us with a new and different way of interpreting the TOV equation.

Specifically, let us now consider two arbitrarily specified functions $\rho(r)$ and $p_r(r)$, and define
\begin{equation}
m(r) = \int_0^r 4\pi \rho(r) r^2 \d r.
\end{equation}
Now consider the line element
\begin{eqnarray}
\d s^2 &=& - \exp\left\{ -2 \int_r^\infty {4\pi[\rho(r)+p_r(r)] r\over1-2m(r)/r} \d r \right\} 
\nonumber
\\
&&
\qquad \times
\left[1-{2m(r)\over r}\right] \d t^2
\nonumber
\\
&&
+ {\d r^2\over 1-2m(r)/r} + r^2 (\d\theta^2 + \sin^2\theta\,\d\phi^2).
\end{eqnarray}
That is, the line element is specified completely by a double iterated integral involving the arbitrary functions $\{\rho(r), p_r(r)\}$. 
Then two of the Einstein equations are trivial, they are ``built in'' to the chosen line element:
\begin{equation}
G_{\hat t \hat t} \equiv 8\pi G_N\, \rho;
\label{E:trivial-1}
\end{equation}
\begin{equation}
G_{\hat r\hat r} \equiv 8\pi G_N\, p_r.
\label{E:trivial-2}
\end{equation}
These two equations permit us to physically interpret $\rho(r)$ as the density and $p_r(r)$ as the radial pressure. 

The sole remaining Einstein equation depends on the $G_{\hat \theta\hat \theta} = G_{\hat \phi\hat \phi}$ component of the Einstein tensor. Now, it is a purely geometrical statement that
\begin{eqnarray}
G_{\hat \theta\hat \theta} &=& G_{\hat \phi\hat \phi} 
\\
&=& G_{\hat r\hat r} +   
4\pi r \left\{ {\d p_r\over \d r} + {(\rho+p_r)(m+4\pi p_r r^3)\over r^2(1-2m/r)} \right\}.
\nonumber
\end{eqnarray}
If we enforce isotropy, $G_{\hat \theta\hat \theta} = G_{\hat \phi\hat \phi} =  G_{\hat r\hat r} $, then without further ado we deduce the TOV equation 
\begin{equation}
{\d p_r\over \d r} = -  {(\rho+p_r)(m+4\pi p_r r^3)\over r^2(1-2m/r)} ,
\end{equation}
which we now see can be interpreted as the integrability condition allowing us to solve the two equations (\ref{E:trivial-1})--(\ref{E:trivial-2}) for a perfect fluid source. Of course once we have the TOV equation, however derived, we can apply the Riccati equation analysis developed earlier in this article.

A bonus of setting things up this way is that it is immediately clear what happens for anisotropic fluids. The third Einstein equation now yields
\begin{equation}
p_t = p_r +  {r\over2} \left\{ {\d p_r\over \d r} + {(\rho+p_r)(m+4\pi p_r r^3)\over r^2(1-2m/r)} \right\}.
\end{equation}
This can either be viewed as a way of calculating the transverse pressure $p_t$ in an anisotropic fluid, or can be rearranged to yield the well-known anisotropic TOV equation
\begin{equation}
{\d p_r\over \d r} = -  {(\rho+p)(m+4\pi p r^3)\over r^2(1-2m/r)} + {2(p_t-p_r)\over r}.
\end{equation}

\section{Discussion}

The purpose of this article has been to develop several ``physically
clean'' solution generating theorems for the TOV equation --- where by
``physically clean'' we mean that it is relatively easy to understand
what happens to the pressure and density profiles, especially in the
vicinity of the stellar core. In many ways this article serves as a
companion paper to reference~\cite{Petarpa}, where related results
were derived in terms of the spacetime geometry.  These new results
refine our previous results, in the sense that it is now clearer when
a mathematical solution of the isotropy equations are ``physically
reasonable'', at least in the region of the stellar core. 

Indeed one
important message is that in theorems {\bf T2}, {\bf T3}, and {\bf T4} of
reference~\cite{Petarpa} one should take $\sigma\neq 0$ if one wishes
the pressure and density to remain well behaved at the center the of
the fluid sphere.

Reference~\cite{Petarpa} also contains a number of other interesting
results regarding the use of ``solution generating'' theorems to
classify perfect fluid spheres --- sometimes new solutions are
obtained, sometimes old solutions are recovered in a new context, and
theorems {\bf T1}--{\bf T4} can be used to generate a web of
interconnections between known and new perfect fluid spheres.  These
comments also apply \emph{mutatis mutandis} to the present theorems,
and we direct interested readers to reference~\cite{Petarpa} for
further details.

In closing, we reiterate that the general relativistic static perfect
fluid sphere, despite its venerable history, continues to provide
interesting surprises.

\section*{Acknowledgements}

This research was supported by the Marsden Fund administered by the
Royal Society of New Zealand. 

In addition, PB was supported by a Royal
Thai Scholarship, and a Victoria University Small Research Grant.  SW
was supported by the Marsden Fund, by a Victoria University PhD
Completion Scholarship, and a Victoria University Small Research
Grant.

\bigskip

\begin{widetext}

\appendix
\section{Key results on Riccati equations}
\def\a{\alpha}
\def\b{\beta}
\def\c{\gamma}

The general Riccati equation is 
\begin{equation}
{\d p(r)\over \d r} = \a(r) + \b(r) \; p(r) + \c(r) \; p(r)^2,
\end{equation}
for specified functions $\a(r)$, $\b(r)$, $\c(r)$. While no completely
general solution from first principles exists, if we are given one
specific solution $p_0(r)$ then the one-parameter general solution (parameterized by an arbitrary real number $k$) may be written~\cite{Bender, Polyanin}:
\begin{equation}
\label{E:R1}
p(r) = p_0(r) +  {k\; \exp\left\{\sint [2\c(r) p_0(r) + \b(r)] \d r\right\}  \over 
1  - k \; \sint \c(r) \exp\left\{\sint [2\c(r) p_0(r) + \b(r)] \d r  \right\} \d r}.
\end{equation}
This can easily be derived, for instance, from the argument sketched
in the standard reference Bender \& Orzag~\cite{Bender}, alternatively
an equivalent explicit statement be found in the reference handbook by
Polyanin and Zaitsev~\cite{Polyanin}.
A second useful result is that if we know two specific solutions
$p_1(r)$ and $p_2(r)$ then the one-parameter general solution (now parameterized by an arbitrary real number $\lambda$) may be written~\cite{Polyanin}:
\begin{equation}
\label{E:R2}
p(r) = 
{\lambda \; \exp\left\{-\sint \c(r) p_1(r) \d r \right\} p_1(r) 
+ (1-\lambda) \; \exp\left\{-\sint \c(r) p_2(r) \d r \right\} p_2(r)
\over
\lambda \; \exp\left\{-\sint \c(r) p_1(r) \d r \right\}  
+ (1-\lambda) \; \exp\left\{-\sint \c(r) p_2(r) \d r \right\} }.
\end{equation}
Finally, if we know three specific solutions $p_1(r)$, $p_2(r)$, and
$p_3(r)$ then the one-parameter general solution (again parameterized by an arbitrary real number $\lambda$) may be written~\cite{Polyanin}:
\begin{equation}
\label{E:R3}
p(r) = { \lambda \;p_1(r)\; [p_3(r)-p_2(r)] + (1-\lambda) \;p_2(r)\; [p_3(r)-p_1(r)]
\over
\lambda \;[p_3(r)-p_2(r)] + (1-\lambda) \; [p_3(r)-p_1(r)]}.
\end{equation}
Note that these three forms of the general solution require two, one, or zero integrations respectively, at the cost of needing one, two, or three pre-specified input solutions. 
\end{widetext}



\end{document}